# All-dielectric silicon metalens for two-dimensional particle manipulation in optical tweezers

Teanchai Chantakit,[1,2,3,†] Christian Schlickriede,[3,†] Basudeb Sain,[3] Fabian Meyer,[3] Thomas Weiss,[4] Nattaporn Chattham[1] & Thomas Zentgraf[3,*]

[1]Department of Physics, Faculty of Science, Kasetsart University, Bangkok 10900, Thailand
[2]College of Advanced Manufacturing Innovation, King Mongkut's Institute of Technology Ladkrabang, Bangkok 10520, Thailand
[3]Department of Physics, Paderborn University, Warburger Str. 100, 33098 Paderborn, Germany
[4]4th Physics Institute, University of Stuttgart, Pfaffenwaldring 57, 70569 Stuttgart, Germany
*Corresponding author: thomas.zentgraf@uni-paderborn.de

[†]The first two authors contributed equally to this work.

**Dynamic control of compact chip-scale contactless manipulation of particles for bioscience applications remains a challenging endeavor, which is restrained by the balance between trapping efficiency and scalable apparatus. Metasurfaces offer the implementation of feasible optical tweezers on a planar platform for shaping the exerted optical force by a microscale-integrated device. Here, we design and experimentally demonstrate a highly efficient silicon-based metalens for two-dimensional optical trapping in the near-infrared. Our metalens concept is based on the Pancharatnam-Berry phase, which enables the device for polarization-sensitive particle manipulation. Our optical trapping setup is capable of adjusting the position of both the metasurface lens and the particle chamber freely in three directions, which offers great freedom for optical trap adjustment and alignment. Two-dimensional (2D) particle manipulation is done with a relatively low numerical aperture metalens ($NA_{ML} = 0.6$). We experimentally demonstrate both 2D polarization sensitive drag and drop manipulation of polystyrene particles suspended in water and transfer of angular orbital momentum to these particles with a single tailored beam. Our work may open new possibilities for lab-on-a-chip optical trapping for bioscience applications and micro to nanoscale optical tweezers.**

## 1. INTRODUCTION

Nowadays, optical tweezers - as cutting-edge technology - pave the way to new intriguing application opportunities in the fields of biophotonics and biomedical research, such as studies of cell interaction, embryology, cancer research or molecular motor characterization [1-4]. In this context, optical micromanipulation includes not only trapping by a non-contact force but also single-cell manipulation, alignment, and sorting of mostly micron-sized dielectric particles [5-7]. Furthermore, digital holographic optical tweezers can be used to generate individual traps to transfer orbital or spin angular momentum and enable the particle circulation and spinning [8-10]. To obtain a stable trapping potential, the gradient force of a tightly focused beam must balance the scattering force exerted on the particle. This is typically accomplished with the use of high numerical aperture (NA) lenses and/or microscope objectives. Current research on optical tweezers is directed towards the design flexibility and versatility in the field of applications, which can be greatly enhanced by replacing bulky and expensive optical elements, such as microscope objectives and spatial light modulators, with miniaturized devices in truly compact setups, suitable for integration into lab-on-a-chip systems. Benefitting from high degrees of freedom in phase modulation and manipulation of the focal characteristics, a polarization-sensitive plasmonic metalens was used to replace bulky refractive elements [11, 12]. However, the relatively low diffraction efficiency of the plasmonic metasurfaces, which affects directly the optical trap efficiency, limits their applicability in optical trapping [13-15].

In recent years, all-dielectric metasurfaces made of low loss and high refractive index materials have been introduced [16-19]. Compared to their plasmonic counterparts, they feature higher diffraction efficiency, lower absorption loss, and a larger optical damage threshold making them a suitable candidate for application in optical tweezers [20]. The utilization of a silicon metalens in optical trapping has been shown recently [21]. The work demonstrated optical trapping with a reflection-based silicon metalens in a microfluidic environment. However, such reflection-based focusing elements require a double-pass of the light through the fluidic system, which can increase damage to biological samples and reduce the focal spot quality by additional scattering processes.

Here, we present a versatile optical tweezers setup based on transmission-type all-dielectric silicon metasurface lenses that can not only optically trap microbeads at a fixed position but also optically manipulate them without using traditional optical elements. The shaping of the intensity profile of the trapping beam by adding a spatially-variant phase modulation to the incident beam is based on the Pancharatnam-Berry (PB) phase concept [22]. The abrupt phase change follows for circularly polarized light that is converted to its opposite helicity. This concept enables our device to work either as a convex or concave lens based on the used input circular polarization state [14, 23, 24]. We demonstrate polarization-sensitive two-dimensional (2D) drag and drop manipulation of polystyrene microbeads suspended in water. Furthermore, we expanded the concept to realize a dielectric vortex metalens, which was used to create a donut-shaped intensity distribution in the focal region without the need for an additional phase mask (q-plate). Theoretical concepts for the orbital angular momentum (OAM) transfer with dielectric vortex metalens already exist but have not been experimentally demonstrated yet [24]. In this work, we show that optically trapped particles can indeed rotate in a circular motion based on the topological charge of the helical phase front. With our approach, we demonstrate metasurface enhanced optical tweezers, which show a high transmission efficiency with simultaneous flexibility in beam shaping that can be used for a broad range of applications in miniaturized 'lab-on-a-chip-ready' systems.

## 2. METHODS

*2.1 Schematic Concept, Metasurface Design, and Nanofabrication*

The concept of the metalens optical tweezers is shown schematically in Fig. 1a. An incident right circularly polarized Gaussian beam at 800 nm wavelength is collected by an ultrathin planar metalens and converted to a left circularly polarized beam that is focused at the designed focal length. Lateral 2D optical trapping of the polystyrene microbeads, which have a refractive index higher than the refractive index of water, near the focus can be described by the momentum conservation of the photons and the beads. The deflected part of the beam from the microbead results in a change of the initial momentum direction and, therefore, in a momentum difference, which implies a net force directed toward the trap center [25].

To implement the metalens, we designed and fabricated a 2D circular nanofin array made of amorphous silicon. The radially changing rotation angle $\theta(r)$ of the nanofins is determined by the desired PB phase modulation $\phi(r) = 2\sigma\theta(r)$, such that $\theta(r) = \frac{\sigma}{2} k_0 \left(\sqrt{f^2 + r^2} - |f|\right)$, where $\sigma = \pm 1$ stands for left or right circular polarization (LCP or RCP), $k_0 = 2\pi/\lambda$ is the free-space wave vector, $r$ is the distance of the nanofin from the center of the lens and $f$ is the focal length of the metalens [13, 14].

By using rigorous coupled-wave analysis (RCWA) with periodic boundary conditions, we found the optimal structure dimensions for a single nanofin with maximum efficiency of polarization conversion from one circular state of polarization to the other. Details of the design method can be found in earlier work [26, 27]. Accordingly, the nanofin geometries are defined by the length of 200 nm, the width of 120 nm, and the center-to-center spacing of 360 nm (Fig. 1b).

We used three different kinds of all-dielectric silicon metasurfaces for our experimental study. A metalens, a linear phase gradient metasurface, and a vortex metalens were fabricated on a 1.1 mm thick glass substrate using silicon deposition, electron beam patterning, and reactive ion etching [28]. At first, a 600-nm-thick amorphous silicon (a-Si) film was prepared through plasma-enhanced chemical vapor deposition (PECVD). Then poly-methyl-methacrylate (PMMA) resist layer was spin-coated onto the a-Si film and baked on a hot plate at 170°C for 2 min. Next, the nanofin structures were patterned by using standard electron beam lithography (EBL). The sample was then developed in 1:3 Methyl isobutyl ketone (MIBK): Isopropyl alcohol (IPA) solution and washed with IPA before being coated with a 20-nm-thick chromium layer using electron beam evaporation. Thereafter, a lift-off process in acetone was executed to remove the remaining PMMA from the surface. We used inductively coupled plasma reactive ion etching (ICP-RIE) to transfer the structures from the chromium mask to silicon. After dry etching the silicon, a thin layer of chromium mask was left on top of the silicon nanofins and we used a wet etching process to remove completely the residual chromium mask.

*2.2 Optical Characterization of the Metalens and Vortex Metalens*

Figures 1c-d show the impact of the polarization-dependent phase modulation by measuring the beam intensity profile along the propagation direction $z$ for different circularly polarized beams incident on the metalens (see also Fig. 3b). For that, we took snapshots of the transverse intensity profiles in incremental steps of 5 μm over the total distance of 825 μm (for RCP and LCP incident light, respectively). The profiles correspond to axial cross-sections of image stacks obtained from different transverse planes. Nearly identical real and virtual focal spots with an FWHM of 0.9 μm are observed at the designed real and virtual focal planes of $z = f = \pm 530$ μm (Fig. 1e-h). The numerical aperture of the metalens in air is NA ≈ 0.6. The diameter of the metalens is 800 μm.

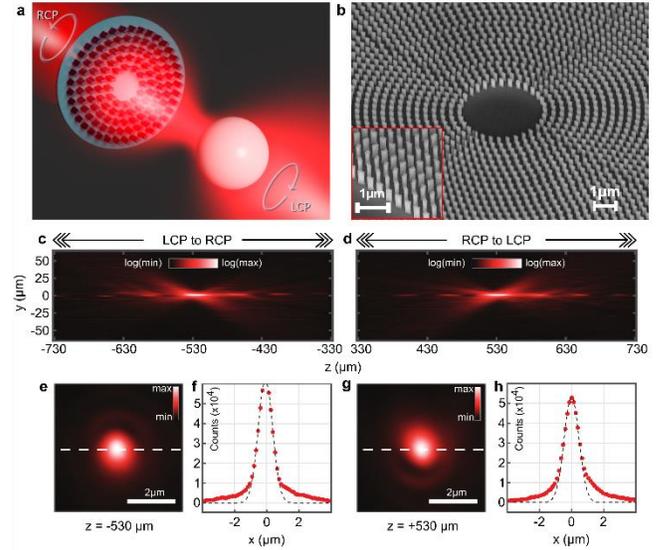

Fig. 1. Schematic concept and optical characterization of the all-dielectric metalens. (a) The conceptual image illustrates the trapping of a polystyrene microbead with the help of an all-dielectric metalens, that converts an RCP incident Gaussian beam to a focused LCP beam. (b) SEM image of the fabricated metalens that consists of amorphous silicon nanofin array. The red inset shows a region at the edge of the metalens. (c) – (d) Cross-sections of the intensity distribution of the focused beam along the optical axis for incident beams with different circular polarizations, drawn on a logarithmic scale. For LCP (RCP) illumination, the metalens acts as a concave (convex) lens, which results in a virtual (real) focal point at $z = -530$ μm ($z = +530$ μm). The metalens is located at $z = 0$. (e) and (g) Transverse intensity distributions at the virtual (real) focal point position, drawn on a linear scale. (f) and (h) Red dots: 1D intensity cross-sections along the white dashed line shown in panel (e) and (g), respectively. Black dashed line: Gaussian fits to the measured data points, which provide an FHWM of 0.9 μm in both considered cases.

The metalens diffraction efficiency is crucial for the application in optical tweezers. Note that any polarization-unconverted light (same polarization as the incident polarization state) does not carry the metalens phase information and therefore is not contributing to the focusing. It only increases the radiation pressure on the particle and decreases the trap efficiency. We measured the metasurface diffraction efficiency by using a metasurface diffraction grating, which is fabricated with silicon nanofin parameters identical to those of metasurface lenses on the same substrate. A conceptual schematic of the measurement is illustrated in Fig. 2a. A focused circularly polarized Gaussian beam incident on the grating is partly converted to the cross-circular polarization at the metasurface position (RCP to LCP or LCP to RCP) and then deflected by the introduced phase grating into the 1st or -1st diffraction order, while the unconverted part of the incident beam causes the 0th diffraction-order in co-circular polarization (RCP to RCP or LCP to LCP). To determine the diffraction efficiency, we measured the k-space intensity distribution for all combinations of input and output circular polarization states (Fig, 2b, see also Fig. 3c for detailed experimental setup). We defined the diffraction efficiency by the ratio of the desired cross-polarized light intensity that is diffracted into the first order to the total amount of light that was transmitted by the metasurface. Diffraction efficiencies of 82.1% for LCP and 83.7% for RCP input light are obtained. From that, we determined the polarization conversion efficiency by multiplying the diffraction efficiency by the transmission coefficient [29]. The transmission coefficient is determined by the ratio of the total intensity transmitted through the metasurface compared to the intensity transmitted through the glass substrate. From these values, we estimate the overall conversion efficiency for the converging metalens and vortex metalens to be 70.8%.

with a helical phase factor $\varphi(x, y) = m \cdot \arctan(x/y)$ that generates a high-quality donut-shaped intensity distribution with a topological charge of $m = \pm 4$, whereas $x$ and $y$ are the center coordinates of each nanofin (Fig. 2e-h) [24]. Depending on the input circular polarization, the phase modulation of the vortex metalens is reversed resulting in either a real focusing vortex beam with a topological charge of $m = +4$ for an incident RCP beam or a virtual focusing vortex beam with inverted OAM helicity ($m = -4$) for an incident LCP beam. The vortex metalens has a diameter of $d_{max} = 400$ µm and a focal length of $f_{ML} = 545$ µm, corresponding to a numerical aperture of NA $\approx 0.35$.

### 2.3 Experimental setup

Next, we characterized the performance of the different fabricated metalenses for optical trapping of microbeads. The metalens optical tweezers setup is shown in Fig. 3a. We used a continuous-wave Ti:sapphire laser at a fixed output wavelength of 800 nm as the light source. The laser power was adjusted with a half-wave plate placed in front of a fixed Glan-Taylor polarizer. By adjusting the quarter-wave plate, we generated different input circular polarizations. The laser beam was weakly focused by a regular convex lens ($f_i$=500 mm) in such a way that the beam waist was slightly larger than the metalens diameter. A non-polarizing 50:50 beam splitter directed both the circularly polarized input beam and the collimated white light illumination onto the sample. The laser beam reflected from the beam splitter was used for power measurement. The metalens focused the RCP beam into the polystyrene microbeads solution contained in a sample chamber formed by a concavity glass slide and a cover glass. Both the metalens and the sample chamber were adjusted freely using independent three-dimensional translation stages. The laser beam profile and the white light image of the microbeads at the same lateral plane were imaged on the CMOS camera (Thorlabs DCC1545M) by a Nikon CFI60 Plan Epi infinity-corrected microscope objective ($\times 100$/ 0.8) and a tube lens ($f_{TL}$=200 mm). To block the laser power and track the particle positions, we used a short-pass filter in front of the camera. The experimental setup can be easily switched to the optical characterization measurements, such as the propagation experiment (Fig. 3b) and the diffraction efficiency measurement (Fig. 3c).

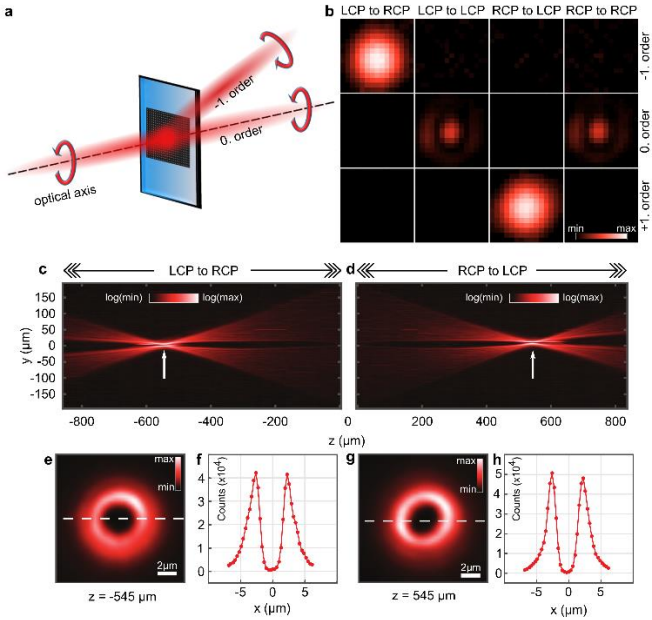

Fig. 2. Measurement of the metasurface diffraction efficiency and the optical characterization of the vortex metalens. (a) Schematic image of the diffraction efficiency measurement: The metasurface diffraction grating with LCP incident light deflects the RCP beam to the -1st order of diffraction, while the unconverted LCP part remains at 0th diffraction order. (b) The cross-section intensity distribution of 1st, 0th and -1st order of diffraction for the incident LCP and RCP beams, further divided into the respective co- and cross-polarization states. (c) and (d) The cross-section intensity distributions of the beams converted by the vortex metalens along the optical axis on a logarithmic scale for better visibility. For LCP to RCP (RCP to LCP) conversion, the metalens acts as a concave (convex) vortex lens, which results in a virtual (real) focal point at $z = -545$ µm ($z = +545$ µm). The helical phase factor results in zero intensity on the optical axis in the focal region. (e) and (g) Transverse intensity distributions of the donut-shaped beam profiles at the focal point positions indicated by white arrows in panels (c), (d), drawn on a linear scale. (f) and (h) Red dots: 1D intensity cross-sections along the white dashed lines shown in panels (e) and (g), respectively.

As a next step, we characterized the optical properties of the vortex metalens in the same way as for the regular metalens (Fig. 2c-h). For fabricating the vortex metalens, we used the metasurface design flexibility to superimpose the parabolic phase profile of the regular lens

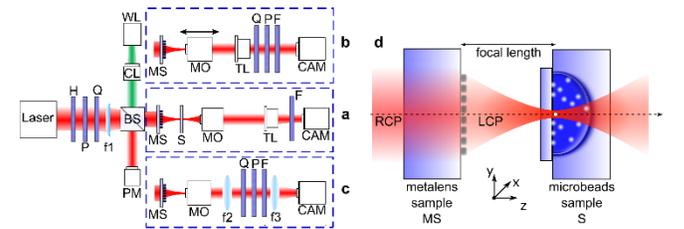

Fig. 3. Schematic illustration of the measurement setup. A non-polarizing beam splitter (BS) is used to insert white light (WL) for sample illumination at the front side of the metasurface (MS - either metalens, vortex metalens, or metasurface diffraction grating) and to measure the incident laser power with a power meter (PM). Three different configurations can be used (blue dashed boxes): (a) Metalens-based optical tweezers system. Laser light with the desired polarization state is focused by the metalens onto the microbeads sample (S) while the focal plane of the metalens is imaged on the camera (CAM) through a microscope objective (MO) and a tube lens (TL). The polarization states were separated by a polarization analyzer consisting of a quarter-wave plate and a linear polarizer. (b) Optical propagation measurement setup. (c) Setup for efficiency measurement with the metasurface diffraction grating. (d) Arrangement of the MS and S in the metalens-

based optical tweezers system. H – half-wave plate; Q – quarter-wave plate; P – linear polarizer; CL – collimating lens; F – filter; $f_{1,2,3}$ – lenses.

## 3. EXPERIMENTAL RESULTS AND DISCUSSIONS

For demonstrating the optical trapping, we dispersed polystyrene microbeads (Polysciences Polybeads) in purified water and loaded the resulting suspension into a concavity glass slide (cavity depth 1.2 - 1.5 mm), which is sealed with a 140-μm thick cover glass. In our experimental setup, the metalens sample and this cover glass were faced towards each other (Fig. 3d). Hence, the working distance of the trapping metalens had to cover a small air gap between the metalens and the microbeads sample covered by the 140 μm thickness of the cover glass. The real focal spot was then generated inside the spherical concavity. This configuration was the reason for working with a relatively large focal length ($f$ > 500 μm), but at the same time, it also offered great flexibility in the trap center adjustment and simple and easy switching between particles of different sizes. To switch to different particles, only the microbeads sample needs to be replaced but the metalens sample remains exactly in the focal plane of the incident beam.

### 3.1 Metalens optical tweezers for 2D particle manipulation

For the measurement, we adjusted the metalens real focal spot in an x-y-plane where polystyrene microbeads were attracted to the cover glass surface. Such surface adhesion forces like van der Waals and electrostatic interaction forces are known as DLVO forces [30]. In a horizontal beam path configuration, the transverse (lateral) gradient force generated by our metalens focus was strong enough to maintain a stable trap in 2D at a laser power of 30 mW. It also stabilized the particles against the force of gravity that tried to pull the particles out of the trap in the y-direction. By tuning the input circular polarization, particles were either be trapped and dragged in the medium (operation as converging metalens) or they were attracted by the surface of the cover glass (operation as concave metalens). Therefore, we can actively tune the 2D gradient force as well as the radiation pressure using the ellipticity of the polarization state. The converging metalens for RCP input generated a focal spot that was smaller than the particle diameter. Therefore, the radiation pressure on the particle increased while it was also partly counteracting the attraction between particle and cover glass. For the 2D lateral trapping, the particle was attracted by the trap center and could be dragged through the solution by moving the microbeads glass slide sample. To drop the particle at the intended location, we had to change the input circular state of polarization, so that the metalens would now work as a concave lens and the beam diverges. Therefore, the radiation pressure on the particle vanished and the particle stuck to the cover glass again. For demonstration purpose, we arranged different lateral pattern in the form of the letters "M", "E", "T" and "A" with different particle diameters ranging from 2.0 to 4.5 μm (Fig. 4a-d). Video files of the M-shaped particle arrangement can be found in the supplementary material (Visualization 1).

We evaluated the lateral trapping stiffness by the standard calibration methods that are based on the particle motion in a stationary optical trap (Fig. 4e) [31, 32]. For that, we used the MATLAB UmUTracker [33] to track particle trajectories of the polystyrene microbeads that freely sank to the bottom in our horizontal optical tweezers setup (particle diameter 4.5 μm). We then compared these results with particles that sank through the 2D optical trap generated by the metalens of NA ≈ 0.6. The velocity with which the particle was attracted to the trap is tracked for different laser powers from 20 to 90 mW. The power $P$ in the trapping plane is reduced due to the metasurface overall efficiency and interface reflections. We found that the maximum lateral trapping forces acting on the bead are lower than $F^{max}$ < 2 pN. The accuracy of the calibration procedure depends on the precision of particle position tracking, which in our system is limited by the magnification of our imaging system and the framerate of the used camera. We assume that the trapping potential is harmonic following a linear dependence of the lateral optical force on the particle distance from the beam axis. The linear fit of the radial stiffness $k_r$ versus power $P$ yields the power normalized radial stiffness of $K_r \approx$ 7.53 pNμm$^{-1}$W$^{-1}$, which agrees with values reported in other works under similar experimental conditions [11]. The trapping efficiency reads $Q_r = k_r r_{max} \frac{c}{nP}$, where $c/n$ is the speed of light in the viscous medium and $r_{max}$ is the maximum displacement. We obtained an efficiency of $Q_r \approx 0.004$. We further confirmed our measurements with radial stiffness simulation based on the Generalized Lorentz-Mie theory (GLMT) [34], which is illustrated in Fig. 4f. As input parameters, we used the focal spot size from the optical characterization measurement that is broadened by a factor of four in the actual metalens optical tweezers system (Fig. 3a), the wavelength of 800 nm, and the focal length of 530 μm. The simulation shows the trap stiffness landscape for different particle diameter and relative refractive indices. For polystyrene particles in water, we find the relative refractive index $\Delta n \approx 1.19$ and for a particle diameter of 4.5 μm the power normalized radial stiffness is $K_r \approx$ 9.5 pNμm$^{-1}$W$^{-1}$. Note that the model predicts that small changes in particle size lead to large changes in trapping stiffness and therefore, simulation and experiment should be compared with caution.

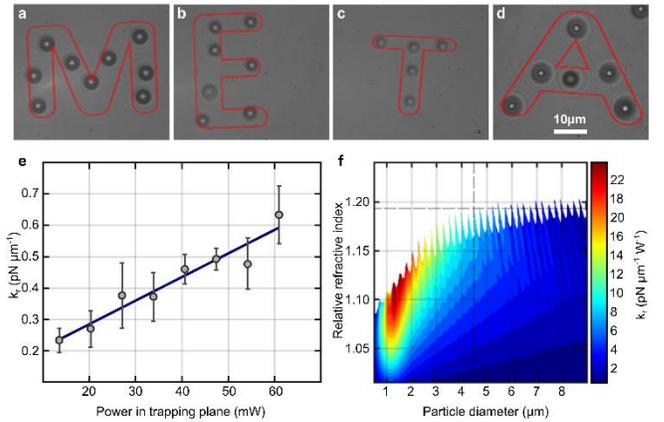

Fig. 4. Metalens for 2D polarization-sensitive drag and drop manipulation of particles. Polystyrene particles with a diameter of (a) 4.5 μm, (b) 3.0 μm, (c) 2.0 μm, and (d) 4.5 μm are dispersed in water and arranged by polarization-sensitive drag and drop using the metalens. (e) Radial stiffness $k_r$ vs power in trapping plane $P$ for polystyrene particles with a diameter of 4.5 μm. (f) GLMT simulation for the radial stiffness of optical trap with different particle refractive indices and various particle diameters. Dashed lines indicate the experimental values of the particle size and the relative refractive index.

### 3.2 Vortex Metalens for OAM transfer with a single tailored beam

In a second metalens optical tweezers experiment, we studied the OAM transfer from our vortex metalens to a polystyrene microbead. We rotated the optical trapping part of the setup (marked with the blue dashed box in Fig. 3a) by 90 degrees so that the gravitational force does not disturb the lateral rotation movement of particles in the x-y-plane (vertical optical tweezers setup with the gravitational force parallel to the z-axis). We only used the sample with polystyrene microbeads of 4.5 μm diameter. The vortex focal spot with the topological charge of $m$ = +4 was adjusted to the lateral region where particles stuck to the cover glass (Fig. 5a). We marked the donut-shaped intensity distribution with red dashed lines and put a short-pass filter in front of the camera to block the laser radiation. We moved the microbeads sample laterally to trap only one particle onto the donut-shaped intensity distribution.

Next, we observed that the polystyrene bead is undergoing a rotational movement along with the vortex beam profile at 19 mW laser power, consistent with the topological charge of the beam (Visualization 2). We tracked the movement in the *x-y*-plane with the help of MATLAB UmUTracker (Fig. 5b). However, the particle was pushed in the axial direction and left the lateral trap potential after approximately 35 seconds. The transient confinement might be caused by inhomogeneous illumination of the vortex metalens, which led to regions on the donut-shaped intensity distribution with lower intensity. After 30 seconds we had to readjust the microscope objective about 12.5 μm in the *z*-direction. Nevertheless, a clear movement along a circular path given by the intensity profile could be observed during that time. Lastly, we compared the power dependence of the radial trap stiffness by repeating the calibration method for the vortex metalens (Fig. 5c). The linear fit now yields a power normalized radial stiffness, which is reduced to $K_r \approx 4.28$ pNμm$^{-1}$W$^{-1}$ and a trapping efficiency of $Q_r \approx 0.002$.

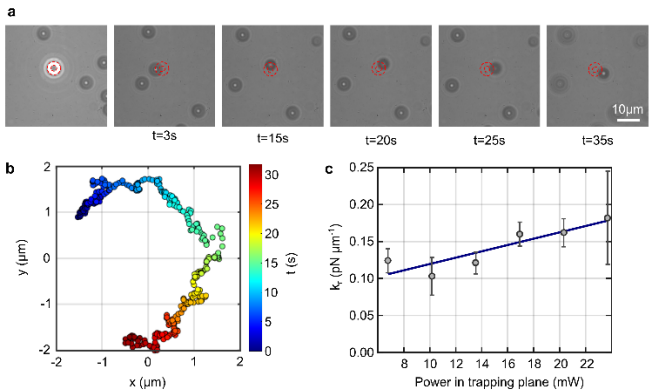

Fig. 5. Vortex metalens for OAM transfer (a) Vortex metalens with a numerical aperture of NA = 0.35 is used to generate a donut-shaped intensity distribution in the focal spot region. At *t* = 0s a particle (4.5 μm diameter) is attracted by the lateral gradient force. The orbital angular momentum is transferred onto the particle resulting in a clockwise rotational movement. Simultaneously, the particle is slowly pushed out of the trap in the axial direction (b) Trajectory plot of the particle's rotational movement. (c) Radial stiffness $k_r$ of the vortex trap versus power in trapping plane *P*.

## 4. CONCLUSION

In summary, we demonstrated efficient all-dielectric transmission-type metasurfaces made of Si nanofins for optical micromanipulation in 2D optical tweezers. We utilized the geometric Pancharatnam-Berry phase to enable a switchable metalens functionality – a convex and a concave lens based on the circular input polarization. With this concept, we could realize a polarization-sensitive drag and drop manipulation of polystyrene microparticles dispersed in water at a power-normalized radial stiffness of $K_r \approx 7.53$ pNμm$^{-1}$W$^{-1}$. Furthermore, we showed the OAM transfer onto particles with the help of a vortex metalens, realizing both vortex beam generation and focusing by one single metasurface element at a radial stiffness per unit power of $K_r \approx 4.28$ pNμm$^{-1}$W$^{-1}$. Hence, no additional phase masks for beam shaping were required. Our work paves the way for future devices based on metalens optical tweezers with possible integration of electronically addressable liquid crystals to switch the polarity of the metalens and that enable fully remotely controlled lab-on-chip optical tweezers.

**Funding Information.** This project has received funding from the European Research Council (ERC) under the European Union's Horizon 2020 research and innovation programme (grant agreement No 724306).

**Acknowledgment.** T.C. acknowledges the support of the Science Achievement Scholarship of Thailand (SAST). The authors acknowledge the continuous support by Cedrik Meier by providing access to the electron beam lithography system.

**References**

[1] Xiao Li, Jun Chen, Zhifang Lin, and Jack Ng. Optical pulling at macroscopic distances. *Science Advances*, 5(3):eaau7814, 2019.

[2] K Dholakia and T Cižmár. Shaping the future of manipulation. *Nature Photonics*, 5(6):335, 2011.

[3] Subra Suresh, J Spatz, JP Mills, Alexandre Micoulet, M Dao, CT Lim, M Beil, and T Seufferlein. Connections between single-cell biomechanics and human disease states: gastrointestinal cancer and malaria. *Acta Biomaterialia*, 1(1):15–30, 2005.

[4] Patrick Lie Johansen, Federico Fenaroli, Lasse Evensen, Gareth Griffiths, and Gerbrand Koster. Optical micromanipulation of nanoparticles and cells inside living zebrafish. *Nature Communications*, 7(1):1–8, 2016.

[5] Pei Yu Chiou, Aaron T Ohta, and Ming C Wu. Massively parallel manipulation of single cells and microparticles using optical images. *Nature*, 436(7049):370, 2005.

[6] David G Grier. A revolution in optical manipulation. *Nature*, 424(6950):810, 2003.

[7] Michael P MacDonald, Gabriel C Spalding, and Kishan Dholakia. Microfluidic sorting in an optical lattice. *Nature*, 426(6965):421, 2003.

[8] V Garces-Chavez, David McGloin, H Melville, Wilson Sibbett, and Kishan Dholakia. Simultaneous micromanipulation in multiple planes using a self-reconstructing light beam. *Nature*, 419(6903):145, 2002.

[9] Hernando Magallanes and Etienne Brasselet. Macroscopic direct observation of optical spin-dependent lateral forces and left-handed torques. *Nature Photonics*, 12(8):461–464, 2018.

[10] Miles Padgett and Richard Bowman. Tweezers with a twist. *Nature Photonics*, 5(6):343, 2011.

[11] Hen Markovich, Ivan I Shishkin, Netta Hendler, and Pavel Ginzburg. Optical manipulation along an optical axis with a polarization sensitive meta-lens. *Nano Letters*, 18(8):5024–5029, 2018.

[12] Satayu Suwannasopon, Fabian Meyer, Christian Schlickriede, Papichaya Chaisakul, Jumras Limtrakul, Thomas Zentgraf, Nattaporn Chattham, et al. Miniaturized metalens based optical tweezers on liquid crystal droplets for lab-on-a-chip optical motors. *Crystals*, 9(10):515, 2019.

[13] Xianzhong Chen, Ming Chen, Muhammad Qasim Mehmood, Dandan Wen, Fuyong Yue, Cheng-Wei Qiu, and Shuang Zhang. Longitudinal multifoci metalens for circularly polarized light. *Advanced Optical Materials*, 3(9):1201–1206, 2015.

[14] Xianzhong Chen, Lingling Huang, Holger Mühlenbernd, Guixin Li, Benfeng Bai, Qiaofeng Tan, Guofan Jin, Cheng-Wei Qiu, Shuang Zhang, and Thomas Zentgraf. Dual-polarity plasmonic metalens for visible light. *Nature Communications*, 3:1198, 2012.

[15] Xingjie Ni, Satoshi Ishii, Alexander V Kildishev, and Vladimir M Shalaev. Ultra-thin, planar, Babinet-inverted plasmonic metalenses. *Light: Science & Applications*, 2(4):e72, 2013.

[16] M Khorasaninejad, W Zhu, and KB Crozier. Efficient polarization beam splitter pixels based on a dielectric metasurface. *Optica*, 2(4):376–382, 2015.

[17] Mohammadreza Khorasaninejad, Wei Ting Chen, Robert C Devlin, Jaewon Oh, Alexander Y Zhu, and Federico Capasso. Metalenses at visible wavelengths: Diffraction-limited focusing and subwavelength resolution imaging. *Science*, 352(6290):1190–1194, 2016.

[18] Jurgen Sautter, Isabelle Staude, Manuel Decker, Evgenia Rusak, Dragomir N Neshev, Igal Brener, and Yuri S Kivshar. Active tuning of all-dielectric metasurfaces. *ACS Nano*, 9(4):4308–4315, 2015.

[19] Ye Feng Yu, Alexander Y Zhu, Ramón Paniagua-Domínguez, Yuan Hsing Fu, Boris Luk'yanchuk, and Arseniy I Kuznetsov. High-transmission dielectric metasurface with $2\pi$ phase control at visible wavelengths. *Laser & Photonics Reviews*, 9(4):412–418, 2015.

[20] Zhe Xu, Wuzhou Song, and Kenneth B Crozier. Optical trapping of nanoparticles using all-silicon nanoantennas. *ACS Photonics*, 5(12):4993–5001, 2018.

[21] Georgiy Tkachenko, Daan Stellinga, Andrei Ruskuc, Mingzhou Chen, Kishan Dholakia, and Thomas F Krauss. Optical trapping with planar silicon metalenses. *Optics Letters*, 43(14):3224–3227, 2018.

[22] Ze'ev Bomzon, Gabriel Biener, Vladimir Kleiner, and Erez Hasman. Space-variant Pancharatnam–Berry phase optical elements with computer-generated subwavelength gratings. *Optics Letters*, 27(13):1141–1143, 2002.

[23] Nanfang Yu and Federico Capasso. Flat optics with designer metasurfaces. *Nature Materials*, 13(2):139–150, 2014.

[24] Yanbao Ma, Guanghao Rui, Bing Gu, and Yiping Cui. Trapping and manipulation of nanoparticles using multifocal optical vortex metalens. *Scientific Reports*, 7(1):14611, 2017.

[25] Arthur Ashkin. Acceleration and trapping of particles by radiation pressure. *Physical Review Letters*, 24(4):156, 1970.

[26] Thomas Weiss. Advanced numerical and semi-analytical scattering matrix calculations for modern nano-optics. Ph.D. thesis, University of Stuttgart, 2011.

[27] Philip Georgi, Marcello Massaro, Kai-Hong Luo, Basudeb Sain, Nicola Montaut, Harald Herrmann, Thomas Weiss, Guixin Li, Christine Silberhorn, and Thomas Zentgraf. Metasurface interferometry toward quantum sensors. *Light: Science & Applications*, 8(1):1–7, 2019.

[28] Bernhard Reineke, Basudeb Sain, Ruizhe Zhao, Luca Carletti, Bingyi Liu, Lingling Huang, Costantino De Angelis, and Thomas Zentgraf. Silicon metasurfaces for third harmonic geometric phase manipulation and multiplexed holography. *Nano Letters*, 19(9):6585–6591, 2019.

[29] Fei Qin, Lu Ding, Lei Zhang, Francesco Monticone, Chan Choy Chum, Jie Deng, Shengtao Mei, Ying Li, Jinghua Teng, Minghui Hong, et al. Hybrid bilayer plasmonic metasurface efficiently manipulates visible light. *Science Advances*, 2(1):e1501168, 2016.

[30] Praneet Prakash and Manoj Varma. Trapping/pinning of colloidal microspheres over glass substrate using surface features. *Scientific Reports*, 7(1):1–10, 2017.

[31] Natan Osterman. Tweezpal–optical tweezers analysis and calibration software. *Computer Physics Communications*, 181(11):1911–1916, 2010.

[32] Nunzia Malagnino, Giuseppe Pesce, Antonio Sasso, and Ennio Arimondo. Measurements of trapping efficiency and stiffness in optical tweezers. *Optics Communications*, 214(1-6):15–24, 2002.

[33] Hanqing Zhang, Tim Stangner, Krister Wiklund, Alvaro Rodriguez, and Magnus Andersson. Umutracker: A versatile matlab program for automated particle tracking of 2d light microscopy or 3d digital holography data. *Computer Physics Communications*, 219:390–399, 2017.

[34] Timo A Nieminen, Vincent LY Loke, Alexander B Stilgoe, Gregor Knöner, Agata M Branczyk, Norman R Heckenberg, and Halina Rubinsztein-Dunlop. Optical tweezers computational toolbox. *Journal of Optics A: Pure and Applied Optics*, 9(8):S196, 2007.